\documentclass[prd,superscriptaddress,amsmath,twocolumn,nofootinbib]{revtex4-2}

\pdfoutput=1
\usepackage{times}

\usepackage{graphicx}
\usepackage{xcolor}

\usepackage{array}
\usepackage{booktabs}
\usepackage{enumitem}
\usepackage{epsf,latexsym,bbm,euscript}

\usepackage{mathrsfs,mathtools,amsmath,amssymb}
\usepackage{tensor}

\makeatletter
\newcommand*{\defeq}{\mathrel{\rlap{%
			\raisebox{0.3ex}{$\m@th\cdot$}}%
		\raisebox{-0.3ex}{$\m@th\cdot$}}%
	=}
\newcommand*{\eqdef}{=\mathrel{\rlap{%
			\raisebox{0.3ex}{$\m@th\cdot$}}%
		\raisebox{-0.3ex}{$\m@th\cdot$}}%
}
\makeatother

\newcommand{\RNum}[1]{\uppercase\expandafter{\romannumeral #1\relax}}

\allowdisplaybreaks
\usepackage{scalerel}
\usepackage{tikz}
\usetikzlibrary{svg.path}
\definecolor{orcidlogocol}{HTML}{A6CE39}
\tikzset{
	orcidlogo/.pic={
		\fill[orcidlogocol] svg{M256,128c0,70.7-57.3,128-128,128C57.3,256,0,198.7,0,128C0,57.3,57.3,0,128,0C198.7,0,256,57.3,256,128z};
		\fill[white] svg{M86.3,186.2H70.9V79.1h15.4v48.4V186.2z}
		svg{M108.9,79.1h41.6c39.6,0,57,28.3,57,53.6c0,27.5-21.5,53.6-56.8,53.6h-41.8V79.1z M124.3,172.4h24.5c34.9,0,42.9-26.5,42.9-39.7c0-21.5-13.7-39.7-43.7-39.7h-23.7V172.4z}
		svg{M88.7,56.8c0,5.5-4.5,10.1-10.1,10.1c-5.6,0-10.1-4.6-10.1-10.1c0-5.6,4.5-10.1,10.1-10.1C84.2,46.7,88.7,51.3,88.7,56.8z};
	}
}
\newcommand\orcidlink[1]{\href{https://orcid.org/#1}{\mbox{\scalerel*{
				\begin{tikzpicture}[yscale=-1,transform shape]
					\pic{orcidlogo};
				\end{tikzpicture}
			}{X}}}}

\usepackage{url,hyperref}
\hypersetup{colorlinks,linkcolor={blue!55!black},citecolor={red!50!black},urlcolor={blue!45!black},breaklinks=true}

 \allowdisplaybreaks
\begin{document}

\title{Black Hole Radiation Sparsity and Bekenstein Entropy Loss in Non-Commutative Schwarzschild Spacetime}

\author{Abdellah Touati\orcidlink{0000-0003-4478-2529}}
\email{touati.abph@gmail.com}
\affiliation{Department of Physics, Faculty of Sciences of Matter, University of Batna-1, Batna 05000, Algeria}
\affiliation{Department of Physics, Faculty of Exact Sciences, University of Bouira, Bouira 10000, Algeria}

\begin{abstract}
	In this paper, we investigate the sparsity of black hole radiation, the Bekenstein entropy loss, and the total particle emission of a Schwarzschild black hole (SBH) within the framework of non-commutative (NC) gauge theory of gravity. First, we provide a brief review of black hole (BH) thermodynamics, computing both the deformed Hawking temperature and entropy. In this geometry, the divergent behavior of the temperature is removed, and a logarithmic correction to the entropy emerges. We then present the NC corrections to the Bekenstein entropy loss of the SBH alongside the total number of emitted particles. Our results show that the total number of emitted particles is proportional to the entropy behavior of the NC SBH, which is consistent with non-thermal radiation. Finally, we analyze the sparsity of Hawking radiation in this geometry, finding that the NC SBH exhibits extremely sparse radiation ($\hat{\eta} \gg 1$), which diverges at the final stage of evaporation as the black hole ceases to radiate.
\end{abstract}

\maketitle

%---------------------------------------------------------------------------------------
\section{Introduction} \label{sec:introduction}
%---------------------------------------------------------------------------------------

Black-hole thermodynamics is one of the most active research areas, attracting significant interest because it provides a bridge between quantum theory and general relativity \cite{hawking3,harms,vaz,haranas2,hansen1,chen1,hansen,jawad1}. The subject originated from early mechanisms attempting to unify quantum field theory and gravity within a semi-classical framework near the black hole event horizon. This unification was pioneered by Hawking \cite{hawking1,hawking2}, who showed that, like a black body, a black hole emits a thermal spectrum of radiation and therefore can evaporate.

However, the semi-classical theory predicts a divergence of the temperature in the final stages of SBH evaporation, which reflects the approach's limitation as an incomplete quantum-gravity (QG) theory. To address this problem, numerous QG-inspired proposals modify black hole thermodynamics via deformed Heisenberg relations, notably the generalized uncertainty principle (GUP) and the extended uncertainty principle (EUP) \cite{lutfugup1,lutfueup3,lutfueup4,lutfueup5,lutfuegup6,gup1,gup2}, which introduce a minimal length that effectively acts as a high-energy cutoff and removes the divergent behavior. Rainbow gravity \cite{rg1,rg2,lutfugr1,lutfugr2}, which is based on modified dispersion relations (MDRs) at Planckian energies, likewise provides a mechanism that regularizes the Hawking temperature in the final stage of evaporation. Non-commutative (NC) geometry \cite{noncommutative1} plays a similar role by predicting a minimal length scale at the Planck length that can eliminate the divergences in final stage of evaporation. Among NC approaches, one commonly replaces the point-like source by a smeared energy distribution, for example Lorentzian \cite{louranziandistrNC2,nctunn3} or Gaussian distributions \cite{nicolini2,gaussiandistrNC2}, or mixtures of these \cite{anand2025}, or encodes non-commutativity through geometric corrections such as Bopp's shift \cite{nozari1,nozari2} and via NC gauge theory of gravity \cite{zaim1,linares1,abdellah2,Hassanabadi1,abdellah4,abdellah5}.

Another important aspect of Hawking radiation is its sparsity, which has recently attracted considerable attention \cite{sparse1,sparse2,alonso3,feng5,sparse4,sparse3}. In parallel, the Bekenstein entropy loss during black hole evaporation and its relation to the total number of emitted particles have been extensively investigated \cite{alonso1,alonso2,alonso3,feng5}.

In this work we extend these investigations by studying both the Bekenstein entropy loss and the sparsity of Hawking radiation within the NC gauge theory of gravity. Motivated by string theory \cite{seiberg1}, this approach posits that spacetime quantization leads naturally to a form of gravity quantization. The NC geometry is based on the commutation relations between spacetime coordinates,
\begin{equation}
	[x^{\mu},x^{\nu}] = i\Theta^{\mu\nu},
\end{equation}
where $\Theta^{\mu\nu}$ is a real antisymmetric matrix. To preserve unitarity \cite{unitarity1,unitarity2} we restrict to space-space non-commutativity ($\Theta^{0i}=0$). The commutation relation modifies the ordinary product between functions $f(x)$ and $g(x)$, which is replaced by the star (Moyal) product ``$\ast$'' defined as
\begin{equation}\label{eq:star}
	(f \ast g)(x) = f(x)\, e^{\frac{i}{2}\Theta^{\mu\nu}\overleftarrow{\partial_{\mu}}\overrightarrow{\partial_{\nu}}}\, g(x).
\end{equation}
Using the gauge-theory formulation of gravity together with the star product and the Seiberg-Witten (SW) map \cite{seiberg1}, we analyze the total number of particles emitted by an NC SBH and the sparsity of its Hawking radiation.

Concretely, we obtain the NC Schwarzschild metric within this framework and compute NC corrections to the Hawking temperature and entropy. Our results show that the NC Bekenstein entropy loss is not constant (unlike the commutative case): it decreases during the final stage of evaporation and reaches zero at a minimum mass where radiation ceases. Corrections to the total number of emitted particles follow a similar pattern to the entropy, which supports the interpretation of non-thermal radiation. We also examine how NC geometry affects the sparsity of Hawking radiation and compare these features to the commutative case.

This paper is organized as follows. Section~\ref{sec:NCSBH} presents the NC Schwarzschild metric corrections derived using the SW map and the star product. Section~\ref{sec:BTNCS} briefly reviews NC corrections to the SBH temperature and entropy. Section~\ref{sec:BELNP} details corrections to the Bekenstein entropy loss and the number of emitted particles. Section~\ref{sec:NCSHR} examines the impact of NC geometry on the sparsity of Hawking radiation. Finally, Section~\ref{sec:concl} contains our concluding remarks.
%---------------------------------------------------------------------------------------%---------------------------------------------------------------------------------------

%---------------------------------------------------------------------------------------
\section{Non-commutative Schwarzschild black hole}\label{sec:NCSBH}
%---------------------------------------------------------------------------------------

We briefly present the formalism of gravity in NC gauge theory \cite{cham1,chai1,mukhe1,Tajron1}, which employs the de Sitter group as a local symmetry \cite{zet1,zet2}. This framework uses tetrad fields as gauge fields, combined with the star product ($\ast$-product) and the SW map, to construct an NC Schwarzschild metric following Ref.~\cite{Tajron1}. 

The deformed gauge fields $\hat{e}^{a}_{\mu}(x,\Theta)$ are expanded as a power series in $\Theta$ up to second order:
\begin{equation}
	\hat{e}^{a}_{\mu}(x,\Theta) = e^{a}_{\mu}(x) - i\Theta^{\nu\rho}e^{a}_{\mu\nu\rho}(x) + \Theta^{\nu\rho}\Theta^{\lambda\tau}e^{a}_{\mu\nu\rho\lambda\tau}(x) + \mathcal{O}(\Theta^{3}),\label{eq:SWM}
\end{equation}
where
\begin{equation}
	e^{a}_{\mu\nu\rho} = \frac{1}{4}\left[\omega^{ac}_{\nu}\partial_{\rho}e^{d}_{\mu} + \left(\partial_{\rho}\omega^{ac}_{\mu} + F^{ac}_{\rho\mu}\right)e^{d}_{\nu}\right]\eta_{cd},
\end{equation}
and
\begin{widetext}
	%\small
	\begin{align}
		e^{a}_{\mu\nu\rho\lambda\tau}&=\frac{1}{16}\Bigg[\bigg(2\{F_{\tau\nu},F_{\mu\rho}\}^{ab}e^{c}_{\lambda}-\omega^{ab}_{\lambda}(D_{\rho}F_{\tau\mu}^{cd}+\partial_{\rho}F_{\tau\mu}^{cd})e^{m}_{\nu}\eta_{dm}-\{\omega_{\nu},(D_{\rho}F_{\tau\mu}+\partial_{\rho}F_{\tau\mu})\}^{ab}e^{c}_{\lambda}-\partial_{\tau}\{\omega_{\nu},(\partial_{\rho}\omega_{\mu}+F_{\rho\mu})\}^{ab}e^{c}_{\lambda}\notag\\
		&-\omega^{ab}_{\lambda}\partial_{\tau}\big(\omega^{cd}_{\nu}\partial_{\rho}e^{m}_{\mu}+\big(\partial_{\rho}\omega_{\mu}^{cd}+F_{\rho\mu}^{cd}\big)e^{m}_{\nu}\big)\eta_{dm}+2\partial_{\nu}\omega_{\lambda}^{ab}\partial_{\rho}\partial_{\tau}e^{c}_{\mu}-2\partial_{\rho}\big(\partial_{\tau}\omega_{\mu}^{ab}+F_{\tau\mu}^{ab}\big)\partial_{\nu}e^{c}_{\lambda}-\{\omega_{\nu},(\partial_{\rho}\omega_{\lambda}+F_{\rho\lambda})\}^{ab}\partial_{\tau}e^{c}_{\mu}\notag\\ 
		&-\big(\partial_{\tau}\omega_{\mu}^{ab}+F^{ab}_{\tau\mu}\big)\big(\omega^{cd}_{\nu}\partial_{\rho}e^{m}_{\lambda}+\big(\partial_{\rho}\omega^{cd}_{\lambda}+F^{cd}_{\rho\lambda}\big)e^{m}_{\nu}\big)\eta_{dm}\bigg)\eta_{cb}-\omega^{ac}_\lambda\omega^{db}_\nu e^f_\rho F^{gm}_{\tau\mu}\eta_{cd}\eta_{fg}\eta_{bm}\Bigg],\label{eq:SWM2}
	\end{align}
	%\normalsize
\end{widetext}
where $e^{a}_{\mu}$ and $\omega^{ab}_{\mu}$ denote the commutative tetrad field and the spin connection. The brackets are defined as
\begin{align}
	\{\alpha,\beta\}^{ab} &= \left(\alpha^{ac}\beta^{db} + \beta^{ac}\alpha^{db}\right)\eta_{cd}, \\
	[\alpha,\beta]^{ab} &= \left(\alpha^{ac}\beta^{db} - \beta^{ac}\alpha^{db}\right)\eta_{cd}, \\
	D_{\mu}F_{\rho\sigma}^{ab} &= \partial_{\mu}F^{ab}_{\rho\sigma} + \left(\omega_{\mu}^{ac}F^{db}_{\rho\sigma} - \omega_{\mu}^{db}F^{ac}_{\rho\sigma}\right)\eta_{cd}.
\end{align}
The inverse tetrad $\hat{e}_{a}^{\mu}$ satisfies the orthonormality relations
\begin{equation}
	\hat{e}_{\mu}^{b} \hat{e}_{a}^{\mu} = \delta_{a}^{b}, \quad \hat{e}_{\mu}^{a} \hat{e}_{a}^{\nu} = \delta_{\mu}^{\nu}.
\end{equation}

The NC metric $\hat{g}_{\mu\nu}$ is derived using \cite{chai1}:
\begin{equation}\label{eq:metric}
	\hat{g}_{\mu\nu} = \frac{1}{2}\left(\hat{e}_{\mu}^{b} \ast \hat{e}_{\nu b} + \hat{e}_{\nu}^{b} \ast \hat{e}_{\mu b}\right).
\end{equation}
We adopt the NC matrix
\begin{equation}
	\Theta^{\mu\nu} = \begin{pmatrix}
		0 & 0 & 0 & 0 \\
		0 & 0 & \Theta & 0 \\
		0 & -\Theta & 0 & 0 \\
		0 & 0 & 0 & 0
	\end{pmatrix}, \quad \mu,\nu = 0,1,2,3.\label{eq:theta}
\end{equation}

The commutative tetrad fields are:
\begin{widetext}
	\begin{align}
		\underline{e}_{\mu }^{0}&=\left(\begin{array}{cccc}\left(1-\frac{2 m}{r}\right)^{\frac{1}{2}}, & 0, & 0, & 0\end{array}\right),\notag \\
		\underline{e}_{\mu }^{1}&=\left(\begin{array}{cccc}0, & \left(1-\frac{2 m}{r}\right)^{-\frac{1}{2}}\sin\theta \cos\phi, & r \cos\theta \cos\phi, & -r \sin\theta \sin\phi\end{array}\right), \notag\\
		\underline{e}_{\mu }^{2}&=\left(\begin{array}{cccc}0, & \left(1-\frac{2 m}{r}\right)^{-\frac{1}{2}}\sin\theta \sin\phi, & r \cos\theta \sin\phi, & r \sin\theta \cos\phi\end{array}\right), \notag\\
		\underline{e}_{\mu }^{3}&=\left(\begin{array}{cccc}0, & \left(1-\frac{2 m}{r}\right)^{-\frac{1}{2}}\cos\theta, & -r \sin\theta, & 0\end{array}\right). \label{eq:tetrad}
	\end{align}
\end{widetext}

Following Ref.~\cite{abdellah1}, we compute the NC metric components $\hat{g}_{\mu\nu}$ for the SBH using Eqs.~\eqref{eq:SWM}, \eqref{eq:tetrad}, and the star product \eqref{eq:star}. The resulting line element becomes:
\begin{equation}
	ds^2=-\hat{f}(r,\Theta)c^2dt^2+\frac{dr^2}{\hat{g}(r,\Theta)}+\hat{g}_{22}(r,\Theta)d\theta^2+\hat{g}_{33}(r,\theta,\Theta)d\phi^2,
\end{equation}

The non-zero components at leading order in $\Theta$ are:
%\small
\begin{widetext}
	\begin{align}
		-\hat{f}(r,\Theta)&=\left(1-\frac{2 m}{r}\right)+\Theta^{2}\left\{\frac{m\left(88m^2+mr\left(-77+15\sqrt{1-\frac{2m}{r}}\right) -8r^2\left(-2+\sqrt{1-\frac{2m}{r}}\right)\right)}{8r^4(-2m+r)}\right\}+\mathcal{O}(\Theta^4),\label{eq:13}\\
		\hat{g}(r,\Theta)&=\left(1-\frac{2 m}{r}\right)-\left(\frac{m \left(48 m^2-4 r^2 \sqrt{1-\frac{2 m}{r}}-55 m r+4 m r \sqrt{1-\frac{2 m}{r}}-4 r^2\right)}{16 r^4 (2 m-r)}\right)\Theta^{2}+\mathcal{O}(\Theta^{4}),\label{eq:14}\\
		\hat{g}_{22}(r,\Theta)&=r^{2}+\Theta^2\left\{\frac{-64 m^3+14 m^2 r \left(6-\sqrt{1-\frac{2 m}{r}}\right)+r^3 \left(5-3 \sqrt{1-\frac{2 m}{r}}\right)-m r^2 \left(36-19 \sqrt{1-\frac{2 m}{r}}\right)}{8 r (r-2 m)^2}\right\}+\mathcal{O}(\Theta^4),\label{eq:15}\\
		\hat{g}_{33}(r,\theta,\Theta)&=r^{2}sin^2(\theta)+\Theta^{2}\left\{\frac{-8 m^3+m^2 r \left(32-6 \sqrt{1-\frac{2 m}{r}}\right)+r^3 \left(5-3 \sqrt{1-\frac{2 m}{r}}\right)+3 m r^2 \left(5 \sqrt{1-\frac{2 m}{r}}-8\right)}{16r(r-2m)^2}\notag\right.\\
		&\left.+\frac{\left(8 m^3+6 m^2 r \left(\sqrt{1-\frac{2 m}{r}}-2\right)+r^3 \left(\sqrt{1-\frac{2 m}{r}}+1\right)+m r^2 \left(2-7 \sqrt{1-\frac{2 m}{r}}\right)\right)cos(2\theta)}{16r(r-2m)^2}\right\}+\mathcal{O}(\Theta^4).\label{eq:16}
	\end{align}
\end{widetext}
\normalsize
The commutative Schwarzschild metric is recovered in the limit $\Theta \rightarrow 0$. 

As a first step, we verify the existence of the NC event horizon graphically by plotting the inverse radial function $\hat{g}(r,\Theta)$:

\begin{figure}[h]
	\centering
	\includegraphics[width=0.48\textwidth]{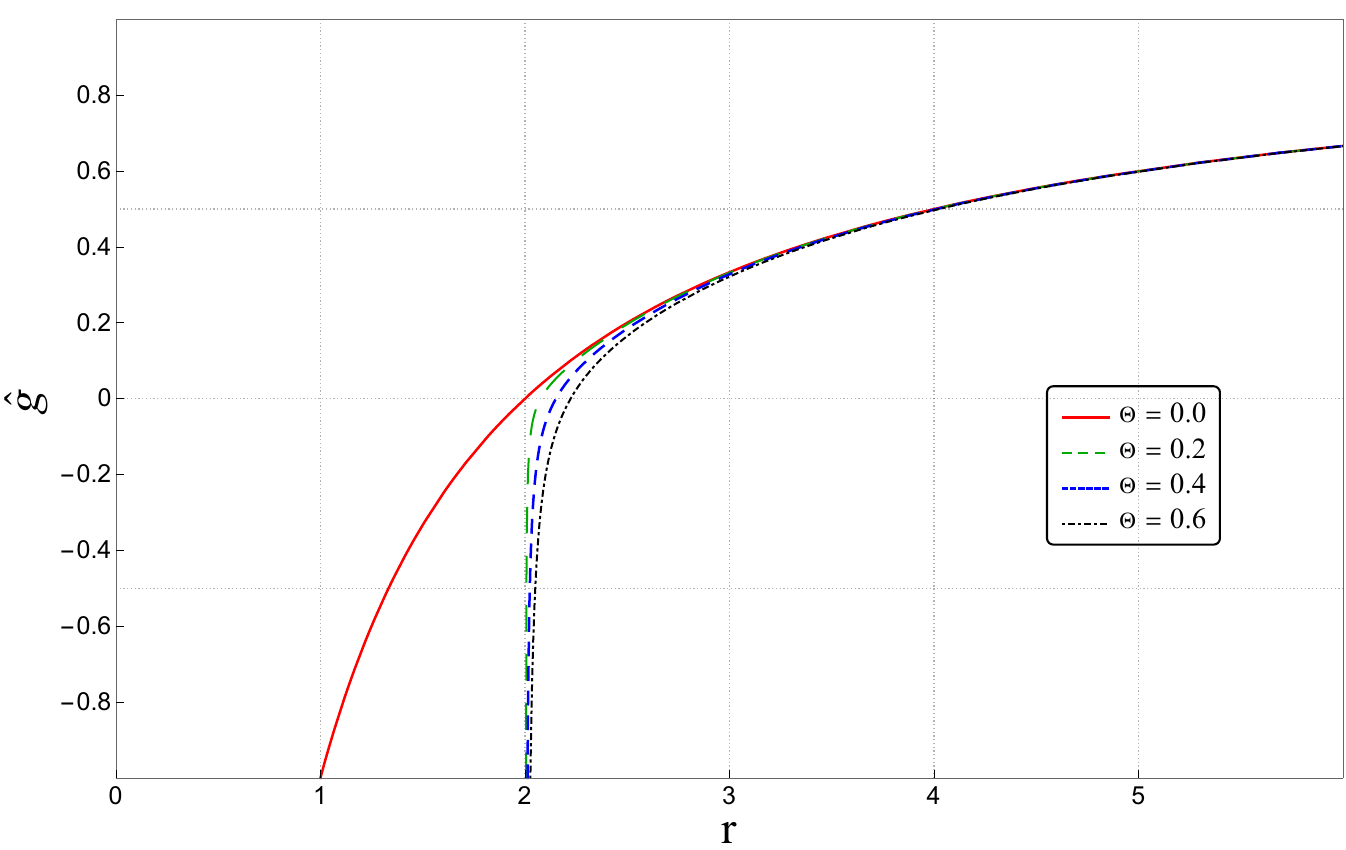}
	\caption{Behavior of the radial component $\hat{g}(r,\Theta)$ as a function of the radial coordinate $r$.}
	\label{fig:grr2}
\end{figure}
As shown in Fig. \ref{fig:grr2}, the radial component $\hat{g}(r,\Theta)$ exhibits a different behavior compared to the commutative one: where the NC event horizon is larger than the commutative one, and the NC singularity is shifted to the finite radius $r=2m$, unlike the model analyzed in Ref. \cite{Tajron1}. 
The NC event horizon is obtained by solving $\hat{g}(r,\Theta) = 0$, which yields (for more details see Appendix~\ref{appA}):
\begin{equation}\label{eq:horizon}
	\hat{r}=2m+\frac{\sqrt{39}}{8}\,\Theta+\frac{\Theta ^{3/2}}{16 \sqrt[4]{39}\sqrt{m}}\, -\frac{23}{128 m}\,\Theta ^2.
\end{equation}
where $m$ denotes the mass of the NC Schwarzschild black hole. The expression for the event horizon reduces to the commutative result in the limit $\Theta\to0$.
%---------------------------------------------------------------------------------------
%---------------------------------------------------------------------------------------

%---------------------------------------------------------------------------------------
\section{Black hole thermodynamics in NC spacetime} \label{sec:BTNCS}
%---------------------------------------------------------------------------------------

In this section, we briefly review some thermodynamic properties of the NC SBH (for more details, see Refs.~\cite{abdellah2,abdellahPhD}). First, we derive the Hawking temperature using the classical definition via the surface gravity $\kappa$. From the deformed metric components \eqref{eq:13} and \eqref{eq:14}, the NC Hawking temperature $\hat{T}_H$ at leading order in $\Theta$ is given by
\begin{align}\label{eq:NCtemperature1}
	\hat{T}_H=\frac{\hat{\kappa}}{2\pi}&=\frac{1}{4\pi\sqrt{\hat{f}(r,\Theta)/\hat{g}(r,\Theta)}}\left.\frac{\partial \hat{f}(r,\Theta)}{\partial r}\right|_{r=\hat{r}},\notag\\
	&=\frac{1}{4\pi \hat{r}}\bigg(1-\frac{\sqrt{39}}{8 \pi }\bigg(\frac{\Theta}{\hat{r}}\bigg)-\frac{1}{8 \sqrt{2} \sqrt[4]{39} \pi }\bigg(\frac{\Theta}{\hat{r}}\bigg)^{3/2}\notag\\
	&-\frac{53 }{128 \pi  }\bigg(\frac{\Theta}{\hat{r}}\bigg) ^2 \bigg),
\end{align}
where $\hat{r}$ is the NC event horizon.

\begin{figure}[htb]
	\centering
	\includegraphics[width=0.48\textwidth]{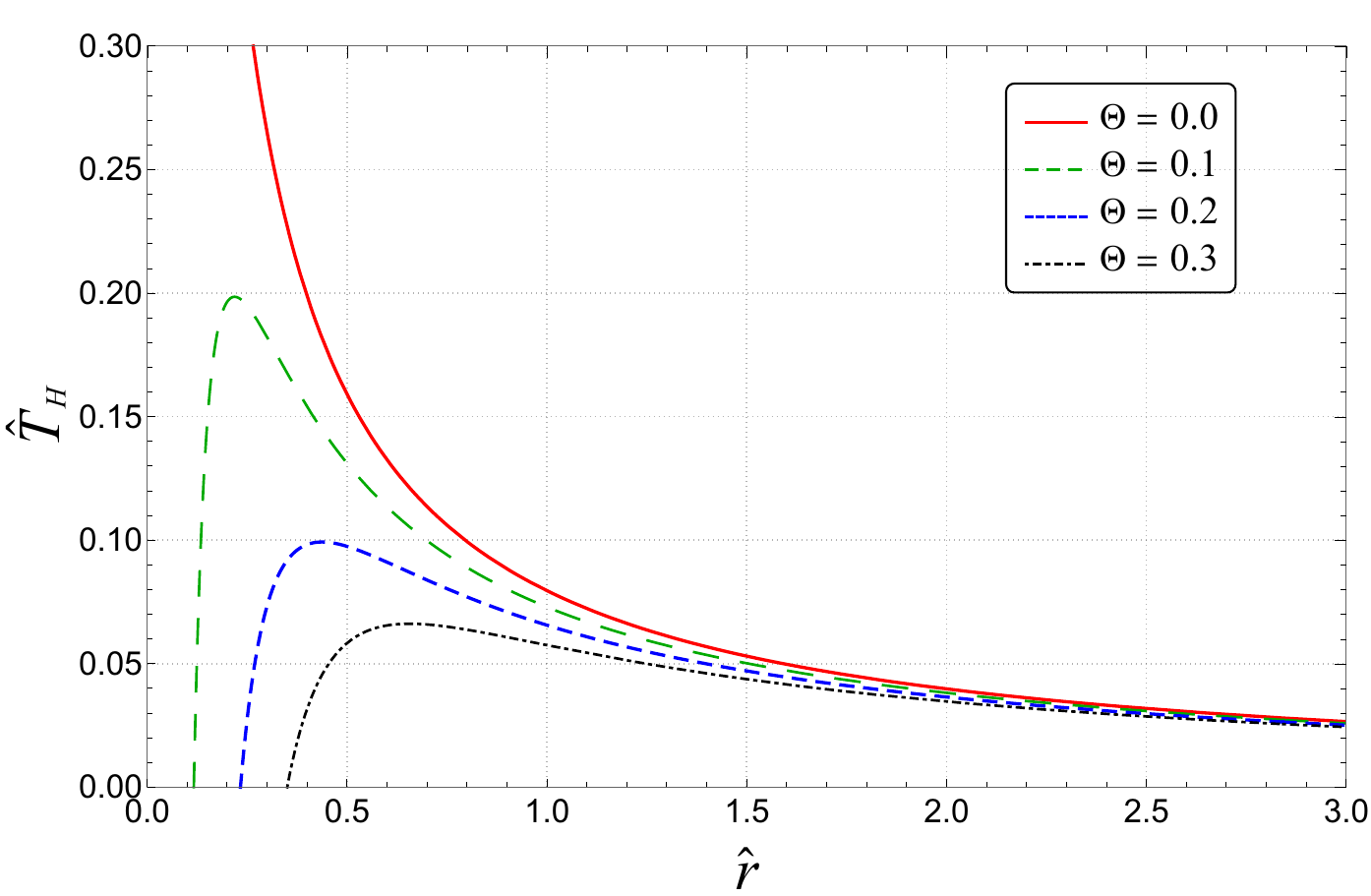}
	\caption{Behavior of the NC Hawking temperature as a function of the horizon radius $\hat{r}$.}
	\label{fig2}
\end{figure}

Figure~\ref{fig2} displays the NC Hawking temperature as a function of the horizon radius $\hat{r}$. As shown, the NC geometry removes the divergence of the Hawking temperature and yields a new maximum temperature $\hat{T}^{\mathrm{max}}_H \approx \frac{0.0199}{\Theta}$ at the critical radius $\hat{r}^{\mathrm{crit}} \approx 2.289\,\Theta$. For $\hat{r}>\hat{r}^{\mathrm{crit}}$ the temperature decreases and eventually vanishes at $\hat{r}^{\mathrm{min}} \approx 1.168\,\Theta$, signaling the end of evaporation.

To estimate the NC parameter we use the back-reaction point $(\hat{T}^{\mathrm{max}}_H,\hat{r}^{\mathrm{crit}})$, following the same steps as Refs.~\cite{abdellah2,abdellah4,abdellah5}. We then find
\begin{equation}
	\Theta\simeq1.1\times 10^{-35}\,\mathrm{m}\sim l_p,
\end{equation}
where $l_p$ is the Planck length. These results suggest that the effects of this geometry become relevant at the Planck scale.

The NC entropy is obtained from the first law of black hole thermodynamics \cite{abdellah4,abdellah5}:
\begin{align}\label{eq:NCentropy1}
	\hat{S}&=\int \frac{dm}{\hat{T}}=\pi \hat{r}^2+\frac{1}{4} \sqrt{39} \pi  \hat{r}\,\Theta+\frac{3^{3/4} \pi }{4 \sqrt{2} \sqrt[4]{13}}\sqrt{\hat{r}}\,\Theta^{3/2}\notag\\
	&+\frac{85\pi}{128}   \Theta^2 \log \left(\pi  r^2\right)\notag\\
	&=S+\frac{\sqrt{39 \pi}}{4}    \sqrt{S}\,\Theta+\frac{3^{3/4} \pi^{3/4} \sqrt[4]{S}}{4 \sqrt{2} \sqrt[4]{13}}\Theta^{3/2}\notag\\
	&+\frac{85\pi  \Theta^2 }{128} \log \left(S\right),
\end{align}
where \(m =\frac{\hat{r}}{2}-\frac{\sqrt{39}}{16}  \Theta  -\frac{ \Theta ^{3/2}}{16 \sqrt{2\hat{r}} \sqrt[4]{39}}+\frac{23 }{128 \hat{r}}\Theta ^2 \) is the NC mass of the SBH (analogous to the Reissner-Nordström mass), and \(S=\pi\hat{r}^2\) is the Bekenstein-like entropy (with \(\hat{r}\) the NC event horizon). The logarithmic correction vanishes for \(\Theta=0\), recovering the area law \(S=\pi r_h^2\). For further discussion of logarithmic corrections in NC gauge theory see Refs.~\cite{abdellah4,abdellah5}.

\begin{figure}[htb]
	\centering
	\includegraphics[width=0.45\textwidth]{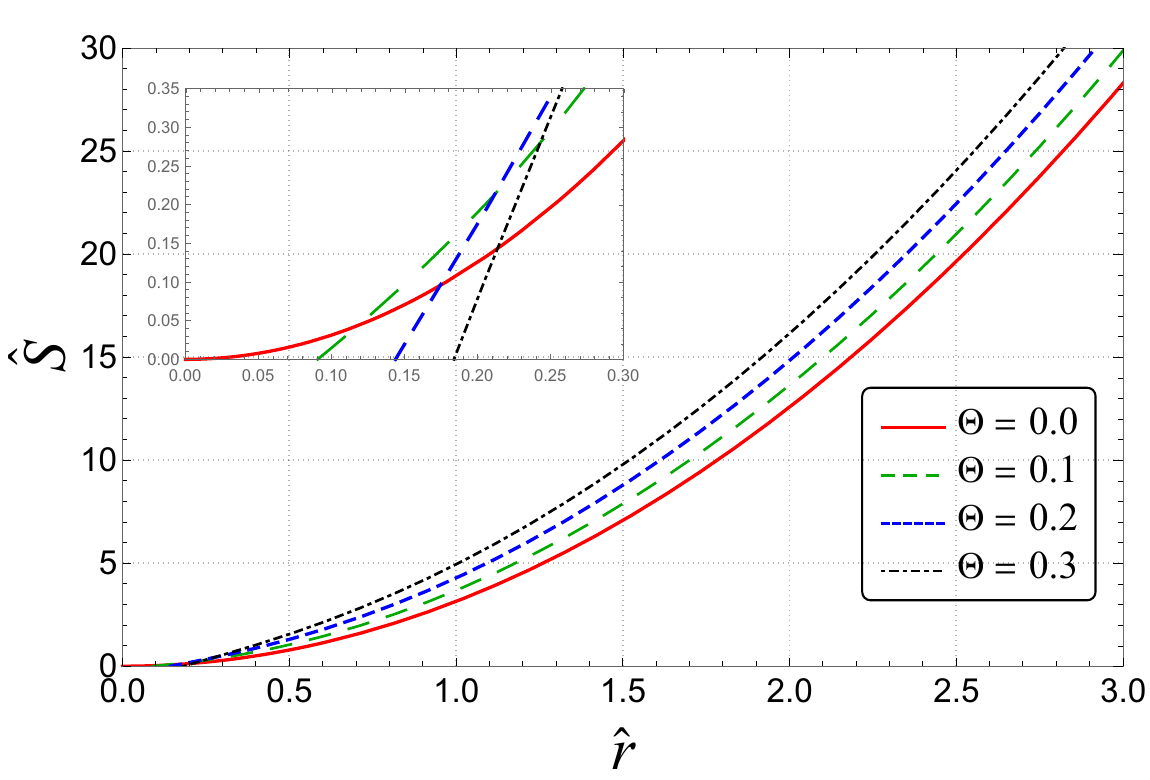}
	\caption{Behavior of the NC entropy as a function of the horizon radius $\hat{r}$.}
	\label{fig:3}
\end{figure}

Figure~\ref{fig:3} shows the NC entropy as a function of \(\hat{r}\) for several values of \(\Theta\). The behavior differs from the deformed area law reported in Ref.~\cite{abdellah2}: non-commutativity increases the entropy for large black holes but decreases it for smaller ones, reaching zero at \(\hat{r}_0\). Moreover, after the black hole stops evaporating (no further thermal radiation is possible), the remnant retains entropy that ultimately reaches zero through a non-thermal process \cite{abdellah4,tunn16,tunn5}.

%---------------------------------------------------------------------------------------
%---------------------------------------------------------------------------------------

%---------------------------------------------------------------------------------------
\section{Bekenstein entropy loss and number of particles emitted} \label{sec:BELNP}
%---------------------------------------------------------------------------------------

In what follows, we investigate the effect of non-commutativity on Bekenstein entropy loss and we check the total number of particles emitted from the NC SBH in the non-thermal regime, following methodologies from other QG models \cite{alonso3,feng5}.

First, the Bekenstein entropy loss per emitted quantum\footnote{In natural units $(G=\hbar=c=1)$.} is given by \cite{alonso2}:
\begin{equation}\label{eq:BEloss}
	\frac{dS}{dN} = \frac{dS/dt}{dN/dt} = 8\pi m \hbar\langle \omega \rangle,
\end{equation}
where $N$ is the total emitted particles and $\langle E\rangle$ is the average energy \cite{alonso1}:
\begin{equation}\label{eq:energycondition}
	\langle E\rangle = \hbar\langle \omega \rangle = \frac{\pi^4}{30\zeta(3)}T_H.
\end{equation}

Using the deformed Hawking temperature \eqref{eq:NCtemperature1}, Equation \eqref{eq:BEloss} becomes:
\begin{align}\label{eq:NCBEloss0}
	\frac{dS}{dN} &=\frac{dS/dt}{dN/dt} = \frac{\pi^4}{30\zeta(3)}\bigg(1 -\frac{\sqrt{39} \Theta}{4 \hat{r}}-\frac{\Theta^{3/2}}{4 \sqrt{2} \sqrt[4]{39} \hat{r}^{3/2}}\notag\\
	&+\frac{71 \Theta^2}{128 \hat{r}^2}\bigg).
\end{align}

From the NC entropy \eqref{eq:NCentropy1}, the Bekenstein entropy loss per quantum for NC SBH is given by:
\begin{align}\label{eq:NCBEloss1}
	\frac{d\hat{S}}{dN} &= \frac{d\hat{S}/dt}{dN/dt} = \frac{dS/dt}{dN/dt}\bigg(1+ \frac{\sqrt{39 \pi } \Theta}{8 \sqrt{S}}+\frac{(3 \pi )^{3/4} \Theta^{3/2}}{16 \sqrt{2} \sqrt[4]{13}\, S^{3/4}}\notag\\
	&+\frac{85 \pi  \Theta^2}{128 S}\bigg).
\end{align}
Substituting \eqref{eq:NCBEloss0} inside the above equation and keeping leading-order $\Theta$ terms:
\begin{equation}\label{eq:NCBEloss2}
	\frac{d\hat{S}}{dN} = \frac{\pi^4}{30\zeta(3)}\left(1-\frac{\sqrt{39} \Theta}{8 \hat{r}}-\frac{\Theta^{3/2}}{16 \sqrt{2} \sqrt[4]{39}\, \hat{r}^{3/2}}\right).
\end{equation}
Noting that, the leading order in the above expression is only the first order, where the second order in canceling during the calculation. It is clear that, the commutative result \cite{alonso2} is recovered for $\Theta \rightarrow 0$.

\begin{figure}[htb]
	\centering
	\includegraphics[clip=true,width=0.48\textwidth]{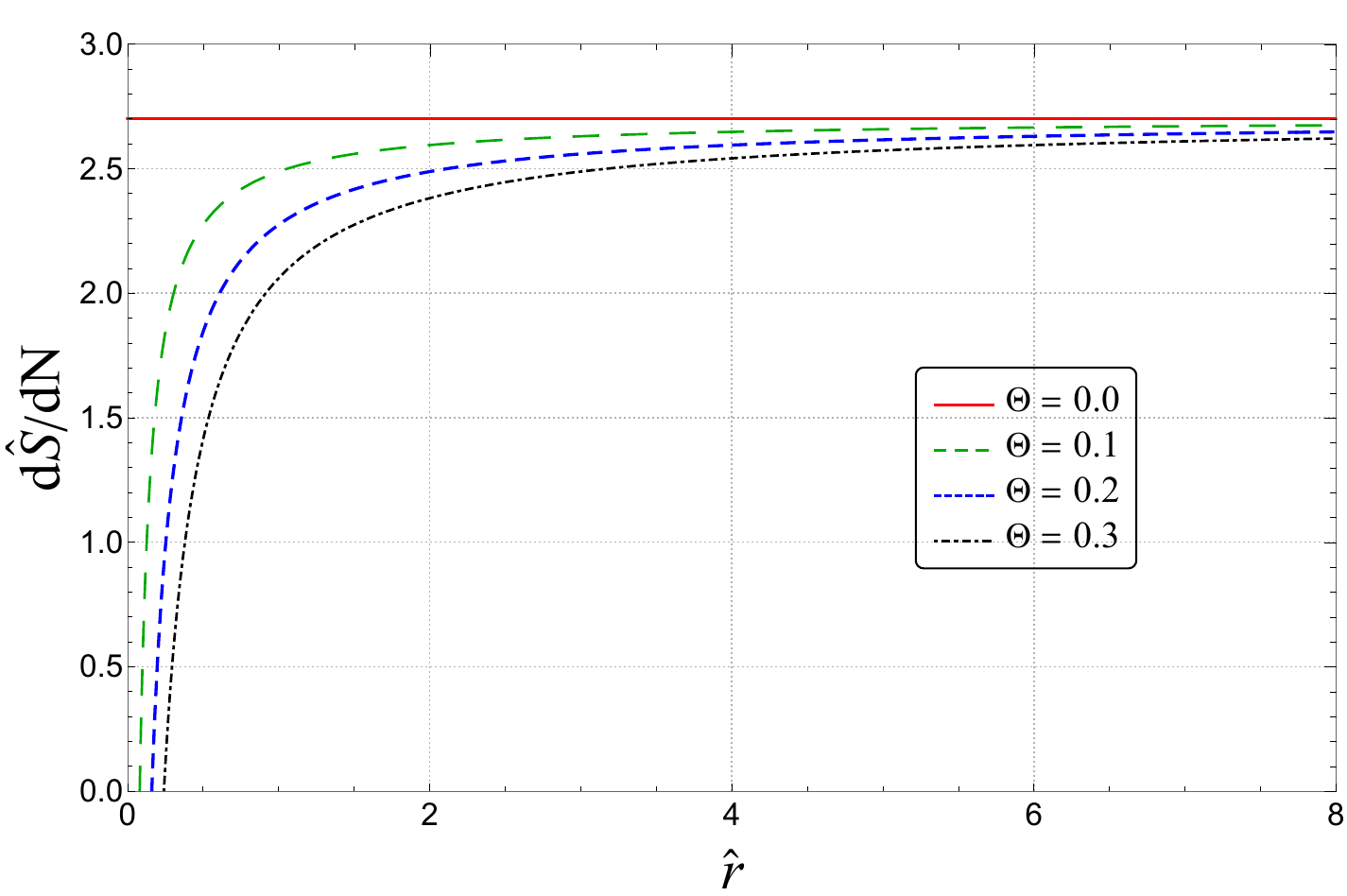}
	\caption{Bekenstein entropy loss per emitted quantum for NC SBH as a function of mass $m$ for varying $\Theta$.}
	\label{fig:NCBEloss1}
\end{figure}
Fig. \ref{fig:NCBEloss1} shows NC Bekenstein entropy loss per emitted quanta  $\frac{d\hat{S}}{dN}$ of NC SBH versus BH event horizon $\hat{r}$. In the commutative case, the Bekenstein entropy loss maintains a constant value of approximately $2.7$ during BH evaporation \cite{alonso1}. For NC scenarios, the effect of this geometry becomes negligible for large BHs (L-BHs) (supermassive ones). However, as evaporation progresses, the quantity $\frac{d\hat{S}}{dN}$ decreases proportionally to the event horizon $\hat{r}$, ultimately reaching zero at a minimum event horizon $\hat{r}_0$. Notably, this behavior aligns with other QG frameworks like RG \cite{feng5,alonso3}, demonstrating similarities between NC effects and QG models, as previously discussed in Ref. \cite{abdellah5}.

Now, we check the effect of the non-commutativity on the total number of particles emitted from NC SBH. For that, we start from the mass element \cite{alonso3}:
\begin{equation}\label{eq:NCtotalN1}
	d\hat{m} = \langle E\rangle d\hat{N}.
\end{equation}
Using Eqs. \eqref{eq:energycondition} and \eqref{eq:NCtemperature1}, we find:
\begin{align}\label{eq:NCtotalN2}
	d\hat{r}&\left(\frac{1}{2} + \frac{\Theta^{3/2}}{32 \sqrt{2} \sqrt[4]{39} \,\hat{r}^{3/2}}-\frac{23 \Theta^2}{128 \hat{r}^2}\right) = \frac{\pi^4 }{30\zeta(3)}\frac{1}{4\pi\hat{r}}\notag\\
	&\times\bigg(1-\frac{\sqrt{39}\Theta}{8 \pi \hat{r}}-\frac{\Theta^{3/2}}{8 \sqrt{2} \sqrt[4]{39} \pi\,\hat{r}^{3/2} }-\frac{53\Theta^2 }{128 \pi \hat{r}^2 }\bigg) d\hat{N}.
\end{align}
The leading-order in $\Theta$ for the element of NC total number of particles $d\hat{N}$ is written as follow:
\begin{align}\label{eq:NCtotalN3}
	d\hat{N} &= \frac{30\zeta(3)}{\pi^4} \bigg(2\pi \hat{r}+\frac{ \sqrt{39} \pi}{4}  \Theta+ \frac{3^{3/4} \pi  \Theta^{3/2}}{8 \sqrt{2} \sqrt[4]{13} \sqrt{\hat{r}}}\notag\\
	&+\frac{85 \pi  \Theta^2}{64 \hat{r}}\bigg)d\hat{r}.
\end{align}
Integrating gives the NC total emitted particles from the NC SBH in the non-thermal radiation scenario:
\begin{align}\label{eq:NCtotalN4}
	\hat{N} &= \frac{30\zeta(3)}{\pi^4}\bigg[\pi \hat{r}^2+\frac{ \sqrt{39} \pi }{4} \hat{r} \Theta +\frac{3^{3/4} \pi  \sqrt{\hat{r}} \Theta^{3/2}}{4 \sqrt{2} \sqrt[4]{13}}\notag\\
	&+\frac{85}{128} \pi  \Theta^2 \log \left(\pi \hat{ r}^2\right)\bigg].
\end{align}
Expressed via commutative entropy $S$:
\begin{align}\label{eq:NCtotalN5}
	\hat{N} &=  \frac{30\zeta(3)}{\pi^4}\bigg[S
	+\frac{\sqrt{39\,\pi}\sqrt{S}}{4}\,\Theta
	+\frac{3^{3/4}\,\pi^{3/4} \sqrt[4]{S}}{4\sqrt{2}\,\sqrt[4]{13}}\,\Theta^{3/2}\notag\\
	&+\frac{85}{128}\,\pi\,\Theta^2 \ln S\bigg].
\end{align}
It is clear that the above expression is proportional to the exact NC entropy expression (see Eq.~\eqref{eq:NCentropy1}), which confirms that, in this case, the radiation of the emitted particles is non-thermal in nature, while for the pure-thermal radiation was obtained in Ref.~\cite{abdellah4}.

\begin{figure}[htb]
	\centering
	\includegraphics[clip=true,width=0.48\textwidth]{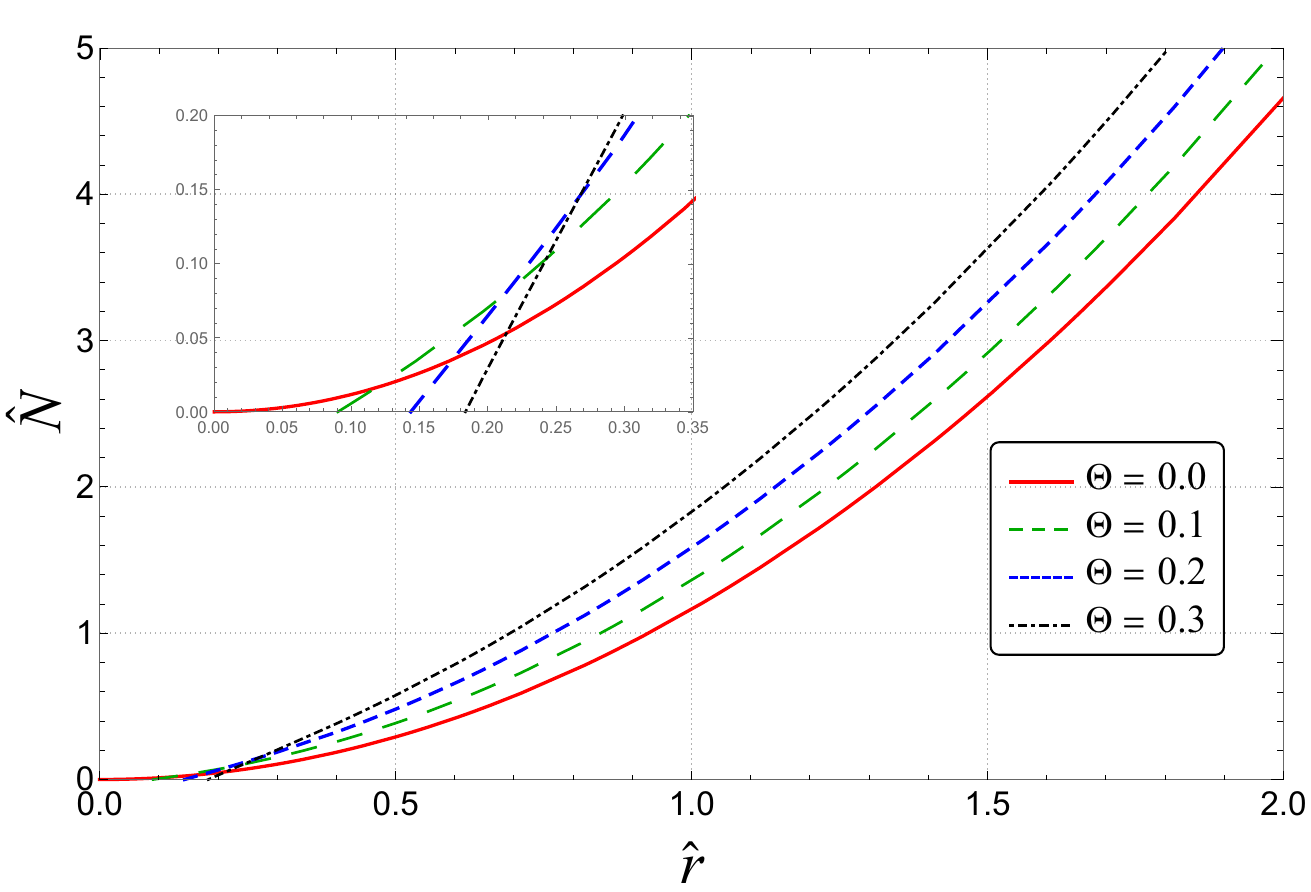}
	\caption{Total particles emitted from NC SBH versus $m$ for varying $\Theta$.}
	\label{fig:NCtotalN}
\end{figure}
The behavior of the NC total number of particles emitted as a function of mass $m$ is shown in Fig.~\ref{fig:NCtotalN}. As observed, this behavior closely resembles that of the NC entropy (see Fig.~\ref{fig:3}), which is expected since the non-thermal radiation spectrum was used to compute the NC correction to the total number of emitted particles. It is evident that the influence of NC geometry in this context differs from the result obtained in Ref.~\cite{abdellah4} for purely thermal radiation. In the pure-thermal case, non-commutativity always reduces the particle emission density, whereas in the non-thermal case, two distinct scenarios emerge. For a large black hole (L-BH), non-commutativity increases the total number of particles emitted, and this increase is enhanced with larger values of $\Theta$. In contrast, for a small BH (S-BH), the effect is reversed: the total number of emitted particles decreases with increasing $\Theta$, until vanishing at the remnant $\hat{r}_{0}$. This behavior again signals the existence of a remnant smaller than the one formed via thermal radiation (see temperature profile Fig. \ref{fig2}), because the non-thermal radiation is allowed when the black hole stop the pure thermal radiation ($\hat{T}_H=0$), and this results due to NC corrections, which does not occur in the commutative case.

%---------------------------------------------------------------------------------------
%---------------------------------------------------------------------------------------

%---------------------------------------------------------------------------------------
\section{Non-commutative correction to the sparsity of hawking radiation} \label{sec:NCSHR}
%---------------------------------------------------------------------------------------

In what follows, we examine an important property of Hawking radiation in the presence of non-commutativity: the sparsity of Hawking radiation. In this context, the sparsity of Hawking radiation is characterized by a dimensionless quantity $\eta$, defined as the ratio between the average time interval between the emission of two successive quanta and the natural time scale \cite{sparse1,sparse2}. This can be expressed as:
\begin{equation}
	\eta = C\frac{\lambda_{T}^2}{gA_{\text{eff}}},
\end{equation}
where $A_{eff}=27A/4$ is the effective horizon area, $\lambda_{T}=2\pi\hbar/(k_B T)$ is the thermal wavelength, and $C$ and $g$ are dimensionless constants representing, respectively, a proportionality factor and the spin degeneracy factor. This parameter helps determine the nature of Hawking radiation for a BH. For $\eta \ll 1$, the BH emits radiation continuously, while for $\eta \gg 1$, the radiation is emitted discretely, indicating that Hawking radiation is extremely sparse.

To obtain the NC effective horizon area $\hat{A}_{eff}=\frac{27A_h^{\mathrm{NC}}}{4}$, we compute the NC horizon area of the NC SBH using the following definition:
\begin{equation}\label{eq:NCarea1}
	A_h^{\mathrm{NC}}=\int_{0}^{2\pi}\int_{0}^{\pi}\sqrt{\hat{g}_{22}\hat{g}_{33}}\,d\theta\, d\phi\,.
\end{equation}

Using the NC deformed metric components \eqref{eq:15}, \eqref{eq:16}, together with the event horizon \eqref{eq:horizon}, we find:
\begin{align}
	\hat{A}&=2\pi\,\int_0^{\pi}\bigg(\hat{r}^2+\bigg[\frac{(4 m+\hat{r}) \big(\hat{r} \csc ^2(\theta ) \left(3 m^2-3 m \hat{r}+\hat{r}^2\right)}{8 \hat{r}^4}\notag\\
	&+\frac{m \left(-26 m^2+25 m \hat{r}-4 \hat{r}^2\right)\big)}{8 \hat{r}^4}\bigg]\bigg)\sin\theta\,d\theta,\notag\\
	&=4\pi\hat{r}^2+\frac{3 \pi}{2}\Theta^2,
\end{align}
where the $\csc$ terms are calculated using the residual amount \cite{Hassanabadi1}.
The NC thermal wavelength is given by:
\begin{align}\label{eq:lambda}
	\hat{\lambda}_{T}&=\frac{2\pi\hbar c}{k_B \hat{T}_H}\notag\\
	&=\frac{2\pi}{ \frac{1}{4\pi \hat{r}}\bigg(1-\frac{\sqrt{39}\Theta}{8 \pi \hat{r}}-\frac{\Theta^{3/2}}{8 \sqrt{2} \sqrt[4]{39} \pi\,\hat{r}^{3/2}}-\frac{53\Theta^2}{128 \pi\,\hat{r}^2} \bigg)}
\end{align}

The NC sparsity of the deformed Hawking radiation is now defined by:
\begin{equation}
	\hat{\eta}=C\frac{\hat{\lambda}_{T}^2}{g\hat{A}_{eff}}
\end{equation}
At leading order in $\Theta$, we obtain:
\begin{align}
	\hat{\eta}&=\frac{C}{g}\bigg(\frac{64\pi^3}{27}+\sqrt{\frac{13}{3}} \frac{16 \pi ^3 }{9 }\bigg(\frac{\Theta}{\hat{r}}\bigg)+\frac{8 \sqrt{2} \pi ^3 }{27 \sqrt[4]{39} }\bigg(\frac{\Theta}{\hat{r}}\bigg)^{3/2}\notag\\
	&+\frac{146 \pi ^3}{27}\bigg(\frac{\Theta}{\hat{r}}\bigg)^2\bigg).
\end{align}

\begin{figure}[htb]
	\centering
	\includegraphics[clip=true,width=0.48\textwidth]{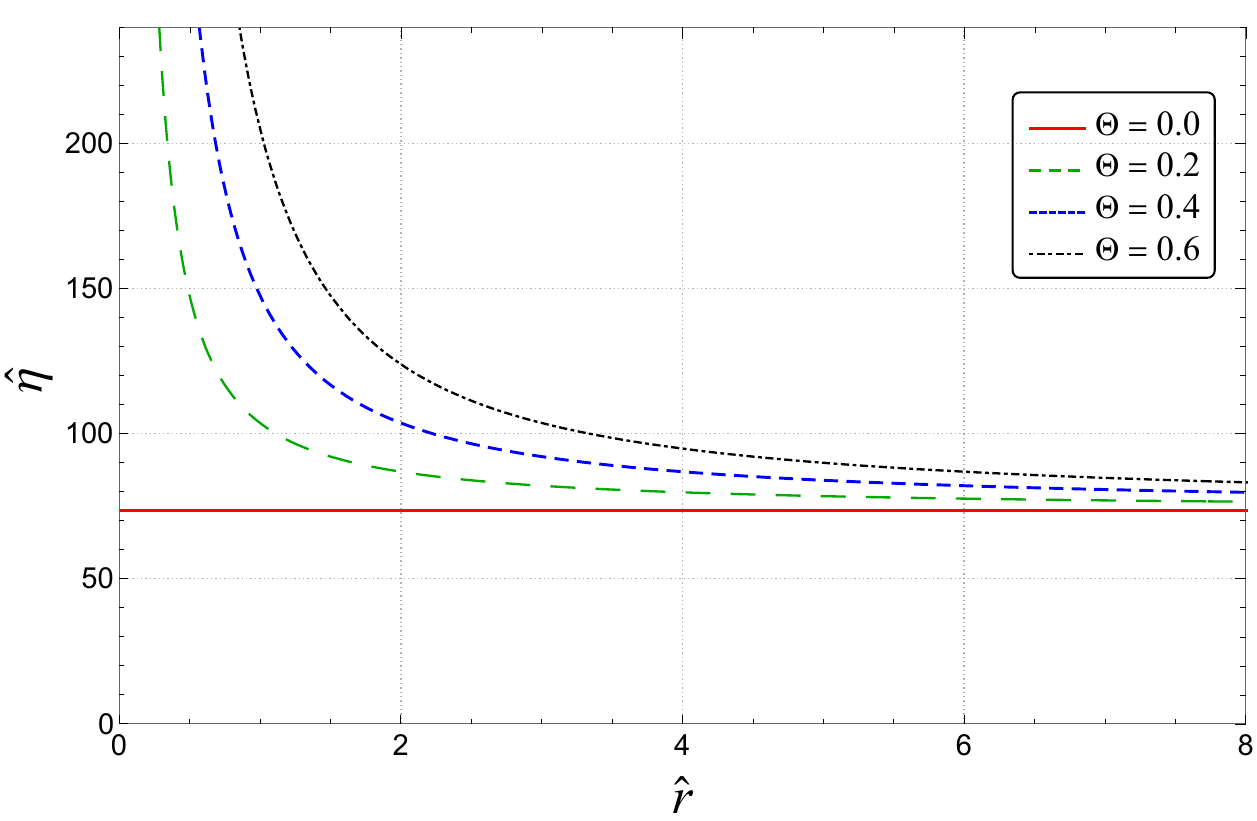}
	\caption{NC sparsity of radiation as a function of BH mass $m$ for varying $\Theta$.}
	\label{fig:NCSHR}
\end{figure}

Fig.~\ref{fig:NCSHR}, shows the NC sparsity of Hawking radiation as a function of NC event horizon $\hat{r}$. It is evident that the commutative sparsity of radiation remains constant ($\eta_0 \propto \frac{64\pi^3}{27}$) during the evaporation of a NC SBH. However, this is not the case in NC spacetime, where the NC sparsity increases at the final stage of evaporation and exhibits a divergent behavior. This indicates that Hawking radiation becomes extremely sparse ($\hat{\eta} \gg 1$), meaning the time interval between successive particle emissions is much larger than the time required to emit an individual Hawking quantum. The divergence at the final stage ($\hat{\eta} \rightarrow \infty$) signals the cessation of the evaporation process\footnote{The mean interval between emissions is related to sparsity by $\Delta \hat{t}=\hat{\eta} \hat{\lambda}_{thermal}/c$, implying that when $\hat{\eta} \rightarrow \infty$, the emission interval also diverges, $\Delta \hat{t} \rightarrow \infty$.}, which is consistent with the $\hat{T}-\hat{r}$ temperature profile shown in Fig.~\ref{fig2}.
%---------------------------------------------------------------------------------------
%---------------------------------------------------------------------------------------

%---------------------------------------------------------------------------------------
\section{Conclusions}\label{sec:concl}
%---------------------------------------------------------------------------------------

Using the NC gauge gravity framework, we construct the NC Schwarzschild metric by employing general tetrad fields along with the SW map and the star product. We then briefly present some thermodynamic properties in this geometry. First, we obtain the NC correction to the Hawking temperature, which removes the divergent behavior at the final stage of evaporation and prevents the SBH from complete evaporation. Additionally, the correction to the entropy shows a logarithmic term, which is consistent with other models of QG. 

Next, we examine the effect of this geometry on the Bekenstein entropy loss per emitted quantum for an NC SBH, which decreases during the BH evaporation process until it reaches zero. This implies that the BH ceases to evaporate. We then investigate the effect of non-commutativity on the total number of particles emitted. Our results show a similarity between the behavior of the total number of particles emitted and the entropy, which is consistent with non-thermal radiation (for details on the pure-thermal radiation case, see Ref.~\cite{abdellah4}).

Finally, we analyze the sparsity of Hawking radiation in the presence of non-commutativity. The results show that, in NC spacetime, Hawking radiation becomes extremely sparse ($\hat{\eta}\gg 1$). For L-BHs, this parameter coincides with the commutative case. However, at the final stage of evaporation, the NC sparsity of radiation exhibits divergent behavior, meaning the time interval between successive particle emissions diverges. At this moment, the SBH stops radiating, which is consistent with our analysis of BH thermodynamics.

\acknowledgments
The author acknowledges the Department of Physics at the University of Batna-1 for providing the academic environment and facilities during the initial stages of this research, which was part of the author's doctoral studies.

\appendix

\section{Non-commutative correction to the black hole event horizon} \label{appA}

As can be seen from the behavior of the function $\hat{g}(r,\Theta)$ in Fig.~\ref{fig:grr2}, it admits a real solution, $\hat{g}(r,\Theta)=0$, at leading order in $\Theta$. However, the presence of the square root makes it difficult to obtain an exact analytical solution. To proceed analytically, we expand the NC event horizon as a power series in the NC parameter:
\begin{equation}
	\hat{r}=2m+C_1\Theta+C_2\Theta^2,\label{eq:EH1}
\end{equation}
where $C_1$ and $C_2$ are the NC coefficients obtained by solving the equation $\hat{g}(r,\Theta)=0$ to first and second order in $\Theta$, respectively. Solving this equation to first order in $\Theta$ gives:
\begin{align}
	C_1=\pm\frac{\sqrt{39}}{8}.
\end{align}
We retain only the positive solution in order to obtain a real solution for $C_2$. Substituting this positive root into Eq.~\eqref{eq:EH1} and then solving $\hat{g}(r,\Theta)=0$ to second order in $\Theta$, we find:
\begin{equation}
	C_2=\frac{\Theta ^{-1/2} }{16 \sqrt[4]{39}\sqrt{m}}\,-\frac{23}{128 m} .
\end{equation}
Using these coefficients, the NC event horizon is therefore given by:
\begin{equation}
	\hat{r}=2m+\frac{\sqrt{39}}{8}\,\Theta+\frac{1}{16 \sqrt[4]{39}\sqrt{m}}\,\Theta ^{3/2} -\frac{23}{128 m}\,\Theta ^2.\label{eq:EH2}
\end{equation}
It can be explicitly verified that this analytical expression satisfies the condition $\hat{g}(r=\hat{r})=0$ up to leading order, namely $\mathcal{O}(\Theta^2)$.

%To estimate the NC parameter $\Theta$, we use the requirement that NC corrections must be smaller than experimental measurement accuracy \cite{karimabadi1}. For a primordial BH with mass $m = GM \sim 5 \times 10^{-4} \, \text{m}$ and size $r \sim 1.5 \times 10^{-3} \, \text{m}$ at the final stage of inflation \cite{karimabadi1}, and given astrophysical measurement accuracies of $\sim 1\%$--$10\%$ \cite{ehc,lisa1,fang1}, the upper limit of $\Theta$ is:
%\begin{equation}
%	\Bigg|\frac{T_H - \hat{T}_H}{T_H}\Bigg| = \frac{29\Theta^2}{32m^2} \leq 0.01.
%\end{equation}
%To obtain a physical bound on $\Theta$, we multiply by the square of the cosmological scale factor $a = 10^{-29}$, which ensures physical distances after inflation \cite{PLANCK1}. This follows from our choice of the space-space NC matrix. This yields:
%\begin{equation}
%	\Theta^{\text{Phy}} = \sqrt{a^2\Theta^2} \leq 5.25 \times 10^{-35}\,\text{m} \sim 3.25 l_p,
%\end{equation}

\bibliography{ref}

\end{document}